\documentstyle[aps,12pt,amssymb]{revtex} 
\tighten 
\begin{document}
%
%ENVIRONMENTS
%
\newenvironment{proof}{{\bf Proof:}}{}
\newtheorem{theorem}{Theorem}[section]

\newtheorem{lemma}[theorem]{Lemma}
\newtheorem{proposition}[theorem]{Proposition}
\newtheorem{corollary}[theorem]{Corollary}
\newtheorem{definition}{Definition}[section]
\newcommand{\be}{\begin{equation}}
\newcommand{\bea}{\begin{eqnarray}}
\newcommand{\ee}{\end{equation}}
\newcommand{\eea}{\end{eqnarray}}
\newcommand{\ethbar}{\bar{\eth}}
\newcommand{\Lambdabar}{\bar{\Lambda}}
\newcommand{\p}{\partial} 
\newcommand{\zetabar}{\bar{\zeta}}
\newcommand{\omegabar}{\overline{\omega}}
\newcommand{\mbar}{\overline{m}}
\newcommand{\etabar}{\overline{\eta}}
\newcommand{\rhobar}{\overline{\rho}}
\newcommand{\sigmabar}{\overline{\sigma}}
\newcommand{\alphabar}{\overline{\alpha}}
\newcommand{\xibar}{\overline{\xi}}
\newcommand{\th}{\theta}
\newcommand{\Ld}{\Lambda\, }
\newcommand{\ld}{\lambda}
\newcommand{\tLd}{\bar{\Lambda}\, }
\newcommand{\Wbar}{\overline{W}}
\newcommand{\debar}{\bar{\delta}}

\newcommand{\om}{\omega}
\newcommand{\ombar}{\bar{\omega}}
\newcommand{\Om}{\Omega}
\newcommand{\tOm}{\tilde\Omega}
\newcommand{\z}{\zeta}
\newcommand{\zbar}{\bar{\zeta}}
\newcommand{\Ldp}{\partial_+\Lambda}
\newcommand{\Ldm}{\partial_-\Lambda}
\newcommand{\tLdo}{\partial_1\bar{\Lambda}}
\newcommand{\tLdp}{\partial_+\bar{\Lambda}}
\newtheorem{remark}{Remark}[section]

\newcommand{\Ldo}{\partial_1\Lambda}
\newcommand{\Ldn}{\partial_0\Lambda}
\newcommand{\varid}{\stackrel{\rm def}{=}}
\newcommand{\k}{\vec{k}}
\newcommand{\kcheck}{\check{k}}
\newcommand{\khat}{\hat{k}}
\newcommand{\mI}{(${m}_{I}$) }
\newcommand{\mII}{(${m}_{II}$) }
\newcommand{\g}{{\bf\mbox{g}}}
\newcommand{\f}{{\bf\mbox{f}}}
\newcommand{\scrip}{\cal I^+} 
\newcommand{\scripm}{\cal I^-}
\newcommand{\scri}{\cal I}
\newcommand{\ihat}{\hat{\imath}}
 \newcommand{\jhat}{\hat{\jmath}}
\newcommand{\Hspace}{${\cal H}$-space }
\newcommand{\lone}{\Lambda_1} 
\newcommand{\lonebar}{\Lambdabar_1}
\newcommand{\etatw}{\tilde \eta} 
\newcommand{\zetatw}{\tilde \zeta}
\newcommand{\bareta}{\bar \eta}
\newcommand{\nn}{\nonumber}
\title{Null Surfaces and the Bach Equations}

\author{  Mirta Iriondo	\and Carlos N. Kozameh \and Alejandra Rojas
	} 
\address{      FaMAF				, 
	       Universidad Nacional de C\'ordoba, 
	       5000 C\'ordoba			, 
	       Argentina			
	} 
      
\maketitle 
\begin{center} Abstract
\end{center}

\begin{abstract}
It is shown that the integrability conditions that arise in the Null
Surface Formulation (NSF) of general relativity  (GR) impose a field
equation on the local null surfaces which is equivalent to the
vanishing of the Bach tensor. This field equation is written explicitly
to second order in a perturbation expansion.

The field equation is further simplified if asymptotic flatness is
imposed on the underlying space-time. The resulting equation determines
the global null surfaces of asymptotically flat, radiative
space-times.  It is also shown that the source term of this equation is
constructed from the free Bondi data at $\scri$.

Possible generalizations of this field equation are analyzed. In
particular we include other field equations for surfaces that have
already appeared in the literature which coincide with ours at a linear
level.  We find that the other equations do not yield null surfaces for
GR.

\end{abstract}
%%%%%%%%%%%%%%%%%%%%%%%%%%%%%

\section{Introduction}
Recently, a new formulation of GR has been developed where, instead of
a metric $g_{ab}(x)$ on the space-time, the basic variables are two
functions defined on the bundle of null directions. Denoting by $x^a$
the points of the space-time, by $(\z, \zbar)$ the parameters on the sphere
of null directions, the basic variables in the so called Null Surface
Formulation (NSF) of GR are:

\begin{itemize}
\item a function $Z(x^a, \z, \zbar)$ such that, for fixed values of $(\z,
\zbar)$, its level surfaces, i.e. $Z = const.$ are characteristic
surfaces on the space-time,
\item another scalar function $\Omega$ that, for fixed $(\z, \zbar)$,
plays the role the conformal factor, i.e., fixes the scale to measure
distances on the space-time.
\end{itemize}

It follows from the above properties that $Z$ yields the conformal
structure of the space-time and thus determines nine out of the ten
components of the metric $g_{ab}(x)$ whereas $\Omega$ fixes the last
component. The scalar $\Omega$, however, plays a more important role in
the dynamics of the theory.

The variables  $Z$ and  $\Omega$ satisfy three differential
equations with the following geometrical interpretation:

\begin{enumerate}

\item A complex first order differential equation for $\Omega$ and $Z$ given
schematically as

$$  \eth \Omega = W[Z] \Omega ,$$
with $\eth$ (essentially $\frac{\partial}{\partial \z}$) the covariant derivative on the sphere of null directions \cite{New-Pen}, \cite{ko-new1} and $W$ a functional
that only depends on $Z$

\item A complex differential equation for $Z$ involving derivatives with
respect to $(x^a, \z, \zbar)$.

\item A real equation for $\Omega$ and $Z$ that adopts the form

$$ \partial_1^2 \Omega = Q[Z] \Omega,$$
with $\partial_1$ a derivative along a null geodesic on the $Z$ = const.
hypersurface and $Q$ another functional that only depends on $Z$.

\end{enumerate}

The first two equations must be satisfied by $Z$ and $\Omega$ if they
are to yield a conformal metric for the space-time.  These equations,
called {\em  metricity conditions\/} \mI and \mII are purely
kinematical, since they are valid for any (conformal) metric with
Lorentzian signature. The last equation, denoted by (E) is equivalent
to the vacuum Einstein equations for a lorentzian metric.

If we interpret (E) as an equation for $\Omega$, then the full dynamics
can be thought as an ODE for a single scalar, rather than the usual ten
equations for a metric tensor. The reader should be reminded, however,
that the cut function $Z$ that enters in (E) cannot be arbitrary since
it must satisfy the metricity conditions. In fact, the three equations
(mI), (mII), and (E) must be solved simultaneously for $\Omega$ and $Z$
to obtain a consistent solution to the dynamical problem.
 
In this paper we reexamine the field equations of the NSF and show
that the dynamics of the conformal structure can be written as a single
equation for the function $Z$.

We first observe that the scalar $\Omega$ satisfies two differential
equations, (E) and \mI, where two functionals of $Z$, $Q$ and $W$,
play the role of source terms. Thus, for a non-trivial solution to exist
integrability conditions must be imposed on $Q$ and $W$.

We then study the integrability conditions
of those equations and show that they 
impose a differential equation on the local null surfaces of the NSF.
The resulting equation, together with the metricity condition \mII,
turns out to be equivalent
to the {\em the Bach equations\/}, for a conformal metric and nicely
tie this formalism with previous known results.

We also observe that the new equation, that will also be called Bach
equation,  and \mII are equations only for $Z$, i.e., they do not
include the conformal factor. Thus, they describe the dynamics
of the conformal structure of the space-time.

By imposing globality conditions on the Bach equation, we obtain an
integrodifferential equation that determines the null
surfaces of asymptotically flat, radiative space-times.  In this
equation the Bondi free data plays the role of a source term and its
solutions have a dual meaning. They represent past null cones from
points at $\scri$ and 2-surfaces at $\scri$ representing the
intersection of null cones from interior points with the null boundary.
These 2-surfaces are called light cone cuts of null infinity or simply
l.c. cuts.

Since the field equations that determine the l.c. cuts are so
complicated one is tempted to ask if it is possible to generalize our
formalism , keeping the kinematical arena provided by $\scri$ but
replacing the Bach equation with a simpler equation  for the
2-surfaces.  It is clear that, the surfaces satisfying the new
equations would not yield conformal vacuum metrics but if they, on the
other hand, satisfy \mII they would be characteristic surfaces of an
underlying metric of the space-time with a given dynamical evolution
due to an effective stress-energy tensor.  In particular, we examine
two possible generalizations of the field equations for the cuts and
show that their solutions are not null surfaces of any metric since
equation \mII is not satisfied.  The lesson being learned here is that
although the metricity conditions are kinematical in nature they play
very important role when trying to describe GR via these non local
variables.

This paper is structured as follows. A brief review of the NSF is
presented in Section II whereas the ``standard'' derivation of the Bach
equations is given in Section III. The integrability conditions of the
NSF equations and a second order perturbation procedure together with
the field equations for the light cone cuts  are presented in Section
IV. Possible generalizations of the light cone equations are given in
Section V. All these alternative models fail to satisfy the
metricity conditions. The main results are summarized and possible
applications are discussed in the Conclusions. Several auxiliary
calculations are presented in the Appendices.

%%%%%%%%%%%%%%%%%%%%%%%%%%%%%%%%%%%%%%%%%
\section{ A brief review of NSF}
%%%%%%%%%%%%%%%%%%%%%%%%%%%%%%%%%%%%%%%
Since a thorough description of the NSF is already given in the
literature \cite{ko-new3},\cite{fri-ko-new1},\cite{fri-ko-new2}, we
will summarize its main results without any derivations and defer the
inquisitive reader to the references.

The formalism first introduces a function $Z(x^a,\z,\zbar)$  with $x^a$
points on the space-time and $(\z,\bar\z)$ parameters on the sphere such
that, for each   $(\z,\bar\z)$, $Z=const$ yields a family of surfaces on
the manifold. We ask if $Z=const$ can be thought of null surfaces of a
conformal metric. In general the answer is no since the equation :

\be
   \hat{g}^{ab}(x) Z_{,a}(x^c,\z,\zbar) Z_{,b}(x^c,\z,\zbar) = 0                           \label{g00}
\ee
is an infinite set of algebraic equations, one for each value of
$(\z,\zbar)$, for the nine coefficients of the  conformal  metric
$\hat{g}^{ab}$.  Therefore, we ask what are the conditions to be imposed on
$Z$ such that a nontrivial $\hat{g}^{ab}(x)$ with Lorentzian signature
exists.  Two types of equations arise in the search for these
conditions; the first type yields the components of the conformal
metric, while the second type identifies necessary conditions to be
satisfied by $Z$ (i.e the metricity conditions \mI and \mII).  The
conditions and metric components can be better expressed if we
introduce a null coordinate system given by a set of 4-scalars
associated with $Z$, namely

\begin{equation}
\th^i(x^a,\z,\zbar)= (\th^0,\th^+,\th^-,\th^1)=(Z,\eth Z,\ethbar Z,\eth\ethbar Z) 
 = (u,w,\bar w,R). \label{thetacoord}
\end{equation}                          
We will refer to the $\th^i$ as the {\em intrinsic
coordinate system associated with the parametrized families of
characteristic surfaces}, labeled by the parameters $(\z,\zbar)$.

Introducing its associated 1-form basis  $\th^i_a=\p_a\th^i$ and dual
vectors basis $\th^a_i$ such that $\th^a_i \th^j_a=\delta^j_i$, we find
that the conditions can be written in terms of two scalars

$$\Ld\equiv\eth^2 Z,\hspace{2cm} \Omega^2 \equiv g^{ab} Z_{,a} \eth\ethbar Z_{,b},$$
as:

\begin{eqnarray*}
\qquad(m_{I})\qquad\eth\Omega&=&{1\over 2}W\Omega,\\
\qquad (m_{II})\qquad\partial_-\Ld&=&{1\over 2}\eth\partial_1\Ld-\left(W+\eth(\ln q)\right)\partial_1\Ld,
\end{eqnarray*}
with $q=( 1-\partial_1\Ld\partial_1\tLd)$, $\partial_i \equiv \frac{\partial}{\partial \theta^i}$, and 
$$
W(1-{1\over 4}\partial_1\Ld\partial_1\tLd)=\partial_+\Ld +{1\over 2}\ethbar\partial_1\Ld-{1 \over2}\eth(\ln q)-{1\over 4}\partial_1\Ld\ethbar(\ln q)+{1\over 2}\Ldo(\bar\Ld_{,-}+{1\over 2}\eth\bar\Ldo).
$$

In the above equations, one is implicitly assuming that the scalars $\Ld$ and
$\Omega$ are functions of $(u,w,\bar w,R, \z, \zbar)$, i.e., one assumes that
$\theta^i$ is a well behaved coordinate system, that the relationship (\ref{thetacoord}) can be inverted to write $x^a$ as a function of $\theta^i$, and that the $x^a$ so obtained has been inserted in the r.h.s. of the defining equations for $\Ld$ and $\Omega$. 

It can also be shown that as a consequence of the metricity conditions \mI and \mII one  constructs a metric $\hat g^{ab}(x)$, i.e., independent of the parameters $(\zeta,\bar\zeta)$, as a functional of $\Ld$ and $\Omega$, (c.f. \cite{fri-ko-new1}). The explicit form of the metric is given as :
$$
\hat g^{ab}= \hat g^{ij}(\Ld,\bar\Ld)\th^a_i\th^b_j,
$$
where  $\hat g^{ij}=\Omega^2 g^{ij}$ and
\be
\label{eq:metric}
(g^{ij})=\left ( \begin{array}{cccc}
                  0&0&0&1\\
                  0&-\Ldo&-1&g^{1+}\\
                  0&-1&-\tLdo&g^{1-}\\
                  1&g^{1+}&g^{1-}&g^{11}
                  \end{array}
         \right ),
\ee
with $g^{1+}=-\frac{1}{2}(\ethbar \Ldo+\Ldo\overline{W})=\overline{g^{1-}}$ and $g^{11}=-2-\frac{1}{2}\ethbar^2\Ldo+\eth\partial_-\tLd+{\cal O}(\Ld^2)$.

So far the description of null surface theory has been completely kinematical. Our variables ($\Omega$ and $\Ld$) must satisfy the  metricity conditions to define a Lorentzian metric. We now address the problem of finding a metric that in addition satisfies the trace free vacuum Einstein equations, i.e. 
 $\hat g_{ab}$  satisfies 
\be
\hat R_{ab}-{1\over 4}\hat R \hat g_{ab}=0.
\label{Rab}
\ee
Contracting this equation with $\th^a_1\th^b_1$ and using the explicit form of the metric in this coordinate system yields:
$$
(\mbox{E})\qquad\partial_1^2\Omega=Q\Omega, 
$$
where 
$$Q=-\frac{1}{q}\partial_1^2\tLd\partial_1^2\Ld-\frac{3}{8q^2}(\partial_1 q)^2+\frac{1}{4q}\partial_1^2q.
$$

Thus this equation together with the metricity conditions \mI and \mII build up  a system of differential equations equivalent to the vacuum Einstein  equations.

It should be mentioned that the $\eth$ operator used above adopts a different form when written in the $\theta^i$ coordinate system. We recall that when applied on a function $f(x^a,\z,\zbar)$ , $\eth$ is the covariant derivative with respect to $\z$. However, if we write $x^a = x^a(\theta^i,\z,\zbar)$, then the $\eth$ operator adopts the form
\be
\label{eq:eth}
\eth = \eth ' + w \partial_0 + \Lambda \partial_+ + R \partial_- +  (\ethbar \Lambda - 2 w) \partial _1,
\ee
where $\eth '$ is the partial covariant derivative, i.e. keeping the $\theta^i$ fixed, and $\partial_i \equiv \frac{\partial}{\partial \theta^i}$.

It follows from this equation that the directional derivatives $\partial_i$ do not commute with the $\eth$ or $\ethbar$ operators. We now write the commutation relations (and corresponding notation) between these derivatives since they will be used throughout this work.

Using (\ref{eq:eth}) and denoting by $\delta_i=[\partial_i,\eth]$ and $\bar\delta_i=[\partial_i,\ethbar]$ (c.f. \cite{fri-ko-new1}) we obtain 

\be
\label{eq:com}
\begin{array}{ll}
\delta_1&=
\partial_-+\partial_1 \Ld \partial_++f_1\partial_1,\\
&\\
\delta_-

&=\partial_- \Ld \partial_++f_-\partial_1,\\
&\\
\delta_+
&=\partial_0-({2\over q}-f_+)\partial_1+\partial_+\Ld\partial_+,\\
&\\
\delta_0
&=f_0\partial_1+\partial_0\Ld\partial_+,
\end{array}
\ee
where  
\be
\label{eq:fi}
\begin{array}{ll}
f_1&=
{1\over q}\left(\Ldp+\ethbar\Ldo+\Ldm\tLdo+(\Ldo\partial_+\tLd+\eth\tLdo+\partial_-\tLd)\Ldo\right)\\
&\\
f_+&=
{1\over q}\left(\ethbar\Ldp+\Ldm\tLdp+(\Ldp\partial_+\tLd+\eth\tLdp+\partial_0\tLd)\Ldo\right)\\
&\\
f_-&=
{1\over q}\left(-2\Ldo+\partial_0\Ld+\ethbar\Ldm+\Ldm\partial_-\tLd+(\Ldm\partial_+\tLd+\eth\partial_-\tLd
)\Ldo\right)\\
&\\
f_0&=
{1\over q}\left(\ethbar\partial_0\Ld+\Ldm\partial_0\tLd+(\partial_0\Ld\partial_+\tLd+\eth\partial_0\tLd
)\Ldo\right).
\end{array}
\ee 

For later use we also compute

\be
[\partial^n_1,\eth]=\sum_{i=0}^{n-1}\partial^{n-1-i}_1[\partial_1,\eth]\partial_1^i \qquad \mbox{for}\quad n\in {\bf N}.
\label{com4}
\ee
and
\bea
&&\begin{array}{ll}
\eta_1:&=[\partial_1,\delta_1]\\
&=\partial_1f_1\partial_1+\partial_1^2 \Ld \partial_+,
\end{array}
\label{eta1}\\
&&\nn\\
&&\begin{array}{ll}
\eta_2:&=[\partial_1,\eta_1]\\
&=\partial^2_1f_1\partial_1+\partial^3_1 \Ld \partial_+
\end{array}
\label{eta2}\\
&&\nn\\
&&\begin{array}{ll}
\eta_3:&=[\partial_1,\eta_2]\\
&=\partial^3_1f_1\partial_1+\partial^4_1 \Ld \partial_+.
\end{array}
\label{eta3}
\eea

%%%%%%%%%%%%%%%%%%%%%%%%%%%%%%%%%%
\section{The Bach Equation}
%%%%%%%%%%%%%%%%%%%%%%%%%%%%%%%%%%%%%

The Bach tensor \cite{bach} 

$$
B_{ab}\equiv\nabla^m\nabla^n C_{mabn}+\frac{1}{2}R^{mn}C_{mabn},
$$
is a trace free, symmetric, two index tensor constructed from the
metric of the space-time with a particular useful feature:  under
rescaling of the metric it transforms with a conformal weight. As we
will see below, the vanishing of the Bach tensor (a conformally
invariant statement) arises as a necessary condition for a space-time to be conformal to an Einstein space-time
\cite{ko-new-tod}.

Consider the conformal transformation
$$
g_{ab}=\Omega^2\hat g_{ab}
\quad \mbox{or equivalently}\quad 
\hat g^{ab}=\Omega^2 g^{ab}
$$
and assume that $\hat g_{ab}$ satisfies the trace free vacuum field equations 
(\ref{Rab}). In terms of $\Omega$ and $g_{ab}$ these equations read
\be
\nabla_a\nabla_b \Omega-{1\over 4}\Delta\Omega g_{ab} + \frac{1}{2} \Omega (R_{ab}-{1\over 4}Rg_{ab}) =0.
\label{Einscon}
\ee
We now ask what conditions should be imposed on $g_{ab}$ so that a
solution of (\ref{Einscon}) exists.  If we consider eq. (\ref{Einscon})
as a second order differential equation for $\Omega$ it is then clear
that, for a non trivial solution to exist, integrability conditions
must be imposed  on the metric $g_{ab}$. Taking the curl of
(\ref{Einscon}) and using the Bianchi identities yields
\be
\Omega\nabla^d C_{abcd}-\nabla^d\Omega C_{abcd}=0.
\label{cond}
\ee

Equation (\ref{cond}) is a new  condition imposed on $\Omega$. Thus, we
have to consider a new system  consisting of equations (\ref{Einscon} )
and  (\ref {cond}).

In order to find the integrability conditions of this  system, we take $\nabla^a$ to equation (\ref{cond})

\be
\nabla^a\Omega\nabla^d C_{abcd}-(\nabla^a\nabla^d\Omega) C_{abcd}+\Omega\nabla^a\nabla^d C_{abcd}-\nabla^d\Omega\nabla^a C_{abcd}=0.
\label{eq:intcond1}
\ee
On the other side, assuming that equation (\ref{cond}) is satisfied and using  the symmetries of the Weyl tensor, we obtain

\begin{eqnarray*}
\Omega\nabla^a\Omega \nabla^d C_{abcd}&=&\nabla^a \Omega \nabla^d \Omega C_{abcd}\\
&=&\Omega \nabla^d\Omega \nabla^a C_{abcd}, 
\end{eqnarray*}
this implies that
$$
\nabla^a\Omega\nabla^d C_{abcd}=\nabla^d\Omega\nabla^a C_{abcd}.
$$
Hence, replacing the last expression in (\ref{eq:intcond1}) and using (\ref{Einscon}), we have
$$
B_{bc}\equiv\nabla^a\nabla^d C_{abcd}+\frac{1}{2}R^{ad}C_{abcd}=0.
$$
Thus, the vanishing of the Bach tensor is necessary condition for a metric to be conformally related to an Einstein metric.

A brief review of the above procedure shows that we have taken the following identity

\be
\nabla^a(\nabla_{[a} (\nabla_{b]}\nabla_c \Omega-{1\over 4}g_{b]c} \Delta\Omega) - \frac{1}{2} C_{abc}{}^d \nabla_d \Omega)=0,
\label{geneq1}
\ee
and used the field equations (\ref{Einscon}) and (\ref{cond}) to get rid
of first and second derivatives of $\Omega$ yielding

\be
\label{bach1}
\Omega B_{bc} = 0.
\ee

We will call the l.h.s. of eq. (\ref{geneq1}) the {\it generating integrability
condition} of the conformal Einstein equations. In general, given a PDE
for a field we define its {\it generating integrability condition}  as
the equation constructed with the minimum number of commutators such
that when the field equations are used, all derivatives of the field
disappear. In particular, substituting eq. (\ref{Einscon}) and
(\ref{cond}) in  eq. (\ref{geneq1}) yields the integrability condition
(\ref{bach1}).

It is worth mentioning that a second integrability condition arises if
we take the curl of (\ref {cond}) \cite{ko-new-tod}.  It can be shown,
however, that for asymptotically flat space-times the vanishing of the
Bach tensor is both necessary and sufficient\cite{mason}. Since our
main motivation is to study this class of space-times we will not
explicitly state the second condition.

%%%%%%%%%%%%%%%%%%%%%%%%%%%%%%%%%%%%%%%%%
\section{The integrability conditions}
%%%%%%%%%%%%%%%%%%%%%%%%%%%%%%%%%%%%%%%%%
\label{sec:II}

As we said before,  equations (E), \mI and \mII are the field equations
of the NSF.  Note that \mI (a complex differential equation) and (E)
(an ODE) are two equations that must be satisfied by a real function
$\Omega $. It is therefore natural to study the integrability
conditions of the following system of equations:

\be
\label{eq:system}
\left .\begin{array}{ll}
(\mbox{E})\qquad\partial_1^2\Omega&=Q\Omega,

\vspace{2mm}

\\

\vspace{2mm}

(m_{I})\qquad\eth\Omega&=\displaystyle{{1\over 2}}W\Omega,\\
(\bar m_I)\qquad\ethbar\Omega&=\displaystyle{{1\over 2}}\overline{W}\Omega. 
\end{array}\right \}
\ee
In the above equations the integrability condition between \mI and ($\bar m_I$) is trivially satisfied because $\ethbar^2\Ld=\eth^2\tLd$ (c.f. \cite{ko-new3}), it remains to study the integrability condition between (E) and \mI (and complex conjugate).

Although in principle the procedure to generate the integrabibility
conditions of (\ref{eq:system}) is straightforward, in practice it
becomes quite cumbersome. To illustrate our approach we first consider
a linearized version of (\ref{eq:system}), namely
\be
\label{eq:system1}
\left .\begin{array}{ll}
(\mbox{E'})\qquad\partial_1^2\Omega&=0,

\vspace{2mm}

\\

\vspace{2mm}

(m_{I})\qquad\eth\Omega&=\displaystyle{{1\over 2}}W_1\,\Omega,\\
(\bar m_I)\qquad\ethbar\Omega&=\displaystyle{{1\over 2}}\overline{W}_1 \,\Omega,
\end{array}\right \}
\ee
with $W_1\equiv\Ldp+{1\over 2}\ethbar\Ldo$. 

Note that we are not writting (\ref{eq:system}) perturbately, i.e, we
are not assuming that $\Omega$ has a prescribed behavior in $\Ld$.  It
is important to remark that if we make a perturbation expansion for
both $\Omega$ and the coefficients we do not obtain the linearized
integrability conditions of (\ref{eq:system}). Rather, $\Omega$ has no
prescribed behaviour on $\Lambda$ and the coefficients of
eq.(\ref{eq:system1}) are the linearized version of the corresponding
coefficients of (\ref{eq:system}).

\begin{proposition}
The generating integrability condition of equations (\ref{eq:system1}) is

$$
2\left( \partial_1\eth+ 3\partial_-\right)[\partial_1^2,\partial_+]\Omega = \frac{1}{4}\Omega \partial_1^5\ethbar^2 \Ld= - 3\Omega B_{11},
$$
where  $B_{11}=B_{ab}\theta_1^a \theta_1^b$. The first equality is obtained via the field equations (\ref{eq:system1}) and the second equality is a straightforward calculation of the linearized Bach tensor in the $\theta^i$ coordinate system. Thus, the integrability condition of (\ref{eq:system1}) is given by

$$
\partial_1^5\ethbar^2 \Ld= 0.
$$

 \end{proposition}

\begin{proof}

The proof is tedious but straightforward. It essentially consists of
applying on $\Omega$  the commutators $[\partial_i\partial_j,\eth]$,
$[\partial_i\partial_j,\ethbar]$, and
$[\partial_i\partial_j,\partial_k]$, for $i=(0,+,-,1)$, to generate an
enlarged system of differential equations such that when taking a
further commutator, the first and second derivatives of $\Omega$
disappear.

The details are given in Appendix \ref{Proof1}.${\bf \Box}$
\end{proof}

\vspace{1cm}
\begin{remark}
It is important to note that the term $\partial_1^5\ethbar^2 \Ld$ in
the above proposition has been obtained via two independent
calculations. On one hand it arises from the generating integrability
condition of eq. (\ref{eq:system1}). On the other hand, it comes from
an explicit calculation of the Bach tensor using a linearized metric in
the $\theta^i$ coordinate system.  This is not a mere coincidence. As
we will see below, we will show that the generating integrability
condition of (\ref{eq:system}) is the Bach tensor.

\end{remark}

As mentioned before, the calculations leading to the integrability condition of the system (\ref{eq:system}) are quite involved but follow the same steps as the calculations done at first order in $\Ld$ (for proof  see Appendix  \ref{Proof2}).

\begin{proposition}
\label{propt}
The generating integrability condition of equations (\ref{eq:system}) is

\be
\label{eq:intcond}
\left(\partial_1\eth+ 3 \partial_- + F \partial_1+ \partial_1 F +{3\over 2}(\ethbar\partial_1\Ld +\overline{W} \partial_1\Ld)\partial_1+3\partial_1\Ld \partial_+\right)[\partial_1^2,\partial_+]\Omega+c.c.= - 3\Omega B_{11}
\ee
where $c.c.$ means complex conjugate, and $F=F[\Lambda]$ is given in (\ref{eq:F}).
\end{proposition}

Comparing the two propositions, we immediately see that extra terms of higher order in $\Ld$ arise in  Proposition \ref{propt}. Those counterterms are needed to cancel the first derivatives of $\Omega$ obtained in the full calculation of the commutator $[\partial_1^2,\partial_+]\Omega$.

%%%%%%%%%%%%%%%%%%%%%%%%%%%%%%%%%%%%%%%%%%%
\subsection{Second order perturbation}
%%%%%%%%%%%%%%%%%%%%%%%%%%%%%%%%%%%%%%%%%%%
Note that in  Proposition \ref{propt} we have not calculated the
explicit form of the integrability condition. Although in principle
this can be done, this calculation is extremely involved and does not
shed extra light on the subject. It is of interest however, to compute
the condition up to second order in $\Ld$ since it yields the first non
trivial contribution to the field equations for our non local
variables.

Consider the l.h.s. of equation (\ref{eq:intcond}) and its complex
conjugate.  Since $[\partial_1^2,\partial_+]\Omega$ is ${\cal O}(\Ld)$,
we write $F$ up to order $\Ld$ as $F=3W_1$.  Thus, to second order in
$\Ld$, equation (\ref{eq:intcond}) can be written as

\be
\label{def:B11}
\bigg (\partial_1\eth+3\delta_1+3\partial_1W_1\bigg )[\partial_1^2,\partial_+]\Omega+c.c.= - 3\Omega B_{11} = 0
\ee

We thus have two {\it independent} methods to compute the integrability
conditions to second order; computing the $[1,1]$ component of the Bach
tensor to second order or using the l.h.s. of (\ref{def:B11}). It
should be mentioned that blindly trying to compute the Bach tensor to
second order in the $\theta^i$ coordinate system quickly exceeds the
algebraic power of MAPLE (or alike). We have instead used the
relationships that link several commutators of $\Omega$ with their
corresponding tensorial counterparts to simplify the intermediate steps
of the calculation. For example, using the relationships between
$[\partial_i\partial_j,\eth]\Omega$ and (\ref{Einscon}) and between
$[\partial_i\partial_j,\partial_k]\Omega$ and (\ref{cond}) we can
write the components of Christoffel symbols, Ricci, and Weyl tensor
that are needed for this computation in terms of $Q$ and $W$. The
detailed calculations are given in \cite{teretesis}.  The
final equation reads

\bea
\partial^5_1\eth^2\tLd&=&{1\over 4} \partial^4_1\bigg(\eth \partial_1 \tLd  W +  \partial_1 \tLd  \eth W+[\delta_1,\eth]\tLd\bigg)+3 \eta_1\partial^2_1 \Wbar +2 \eta_2\partial_1 \Wbar+{1\over 2} \eta_3 \Wbar+\nonumber\\
&&+{1\over 2}\bigg(\partial_1\eth+3\partial_-\bigg)(4\partial_+Q+2\partial_1\ethbar Q+{3\over 2}\partial_1^2 \tLd  \partial^2_1 W-\partial_1 \tLd  \partial^3_1 W)+{3\over 2 }\bigg({1\over 2}\partial_1 W\partial_1^3 \Wbar+
\nonumber\\
&&-{1\over 2 }\partial_+  \partial_1 \Ld \partial^3_1 \Wbar+\partial_1^2\Ld \left (\partial^4_1 \tLd -\partial_0\partial^3_1 \tLd 
+{1\over 2}\partial_+(3\partial^2_1 \Wbar-2\partial^2_1 \bar f_1)\right )\bigg)+\nn\\
&&-\frac{3}{4}(3\partial_1^2\Wbar -2 \partial_1^2\bar f_1)\partial_1^2 W-{3\over 2}\partial_1^3\tLd ({\eth\partial_1^2W\over 2}-\partial_-\partial_1W)+c.c.
\label{eq:bach2}
\eea

For simplicity we have dropped the subindex $1$ on $W$ in the r.h.s. of
(\ref{eq:bach2}) since it can only contain linear terms in $\Ld$. As we
will see in the next subsection, the linearized $W$ vanishes for
asymptotically flat space-times. This gives a considerable
simplification in the resulting equation.

It is worth mentioning that using the generating integrability condition to second order yields the same result.

%%%%%%%%%%%%%%%%%%%%%%%%%%%%%%%%%%%%%%%%%%%%%%%%%
\subsection{Asymptotically Flat Spaces-Times}
%%%%%%%%%%%%%%%%%%%%%%%%%%%%%%%%%%%%%%%%%%%%%%%%%%%%%

As was shown by L. Mason, the vanishing of the Bach tensor is both a
necessary and sufficient integrability condition for the existence of a
solution to eq. (\ref{Einscon}) when the underlying space-time is
asymptotically flat along null directions\cite{mason}. Thus, for those
space-times eq. (\ref{eq:bach2}) yields the field equation for the
non-local variable $Z$ up to second order in $\Ld$. On the other hand,
eq. (\ref{eq:bach2}) as it stands is valid even for metrics that are
not asymptotically flat. It is thus clear that asymptotic flatness
imposes extra conditions on $W$ and/or $Q$ on this equation. We now
derive these conditions.

We first notice that $W$ on the r.h.s. must be known to first order in
$\Ld$ since it is always multiplied by factors of order $\Ld$. To
obtain the specific form of the linearized $W$ compatible with an
asymptotic structure we should look for a pair $(\Omega, \Ld)$ that
satisfies eq. (\ref{eq:system1}) with asymptotically flat boundary
conditions.

Asymptotic flatness imposes the boundary condition $\lim_{R \to
\infty}\Omega=1$ on the conformal factor. It then follows from the
first equation (E') that $\Omega=1 + {\cal O}(\Ld^2)$.  Inserting this
in eq. \mI annihilates the l.h.s. of this equation.  Thus, the
linearized $W$ vanishes for asymptotically flat space-times.

It also follows from eq. (\ref{eq:system}) and from  the proper peeling
behaviour of $\Ld$ (see below) that no restriction is impossed on $Q$
(or $W$ to second order in $\Ld$).

A vanishing $W$ (to first order in $\Ld$) gives a big simplification to (\ref{eq:bach2}). We rewrite this equation as

\be
\label{eq:asymp}
\partial^5_1\ethbar^2\eth^2Z={\cal S}(\theta_i,\zeta,\bar \zeta),
\ee
where the function ${\cal S}$ is considered as a source and is given by 

\begin{eqnarray*}
2{\cal S}(\theta_i,\zeta,\bar \zeta)&=&{1\over 2} \partial^4_1\left(-\partial_-\Ld\partial_+\bar\Lambda+ \partial_1\ethbar\Ld \partial_-\bar\Lambda+ \partial_1\Ld\partial_0\bar\Lambda+( \partial_-\ethbar\Ld-2\partial_1\Ld-\eth  \partial_1\ethbar\Ld)\partial_1\bar\Lambda\right)+\\
&&+\left(\partial_1\eth+3\partial_-\right)(4\partial_+Q+2\partial_1\ethbar Q)+3
\partial^2_1 \Ld\big (\partial^4_1 \tLd-\partial_0\partial_1^3\tLd-\partial_+\partial_1^3\eth\tLd\big)+c.c.
\end{eqnarray*} 

In order to  determine the global null surfaces of asymptotically flat, radiative space-times we  integrate eq. (\ref{eq:asymp}) using asymptotic boundary conditions on $\Ld$. 

\vspace{10mm}

We claim that the final equation reads:
 
\begin{eqnarray}
\label{eq:basymt}
\ethbar^2\eth^2Z&=&\left(\eth^2\bar\sigma_B+2 \eth Z \eth\dot{\bar\sigma}_B+(\eth Z)^2 \ddot{\bar\sigma}_B+\sigma_B\dot{\bar\sigma}_B+c.c.\right)-\int_{-\infty}^Z \dot\sigma_B \dot{\bar\sigma}_B \mbox{d} \mu\nonumber\\
&&-\int^{\infty}_R\hspace{-.2in}...\int^{\infty}_{R_4} {\cal S}(\theta_i,\zeta,\bar\zeta) \mbox{d}R_1...\mbox{d}R_5,
\end{eqnarray} 
with $\sigma_B$ the free Bondi data at $\scri$.

\vspace{5mm}

\begin{proof}
Our goal is to show that $\lim_{R\to \infty}\ethbar^2\Ld$ is finite and it is constructed from the free Bondi data $\sigma_B$. It would then follow from the peeling conditions that $\lim_{R\to \infty}\partial_1^i\ethbar^2\Ld=0$, for $i=1...4.$, thus giving only one ``constant'' of integration to eq. (\ref{eq:asymp}).

We consider at $\scri$ the Bondi coordinates $(u,\zeta,\bar\zeta)$ and
the l.c. cut  given  by $u=Z(x^a,\zeta,\bar \zeta)$, where $x^a$
is an interior point of the space-time.

Bondi coordinates come equipped with a null tetrad $(\hat n^a, \hat
l^a, \hat m^a, \hat {\bar m}^a)$  that at $\scri$ have the following
properties:   $\hat n^a$ is a generator of $\scri$, $\hat m^a$ and
$\hat {\bar m}^a$ are tangent vectors along the Bondi cut $u=const$,
and $\hat l^a$  points along the direction of the null hypersurface
that define the Bondi cuts at $\scri$. Furthermore, the vector $\hat
l^a$ is used to construct the shear associated with the bondi cuts as
$\sigma_B \equiv \hat m^a \hat {\bar m}^b \nabla_a \hat l_b$, i.e., it
contains information about the interior of the space-time.

The l.c. cut $u=Z(x^a,\zeta,\bar \zeta)$, on the other hand, is
constructed from the intersection of the future null cone from $x^a$
with $\scri$. This cut also comes equipped with an associated null
tetrad $( n^a, l^a,  m^a,  {\bar m}^a)$ with the following geometrical
properties: $n^a$ is  chosen so that it becomes a generator of $\scri$,
$m^a$ and $\hat{\bar m}^a$ so that they are tangent to $\scri$. The vector
$l^a$ points along the generators of the null cone from $x^a$. At $\scri$
it yields the shear associated with the l.c. cut, namely 
$\sigma_Z \equiv m^a \bar m^b \nabla_a l_b$.

For every point on the l.c. cut we can write its associated null tetrad
in terms of the Bondi tetrad as

\be
\label{eq:tetrad}
\begin{array}{ll}
  n^a&=\hat n^a ,\\
  m^a&=\hat m^a+\eth Z \hat n^a,\\
  l^a&=\hat l^a+\ethbar Z \hat m^a+\eth Z \hat {\bar m}^a+\ethbar Z\eth Z \hat n^a.
\end{array}
\ee
Defining  $\Psi_2$ on the null cone from $x^a$ as

\be
\hspace{1.2cm} \Psi_2=-{1\over 2}(C_{abcd} l^a n^b l^c n^d-C_{abcd} l^a n^b  m^c {\bar{m}}^d),
\label{eq:psi}
\ee
and using the relationship between the null tetrads given by (\ref{eq:tetrad}), we have at $\scri$
 \be
\label{eq:psi2}
 \Psi_2=\Psi_{2B}-2\eth Z \eth {\dot{\bar\sigma}}_B-(\eth Z)^2 \ddot{\bar\sigma}_B,
\ee
where the dot denotes $\hat n^a\nabla_a\equiv\partial_u$.

Solving  the Bianchi identity corresponding to ${\dot\Psi}_{2B}$, i.e. in the Bondi coordinates, we have 
$$
\Psi_{2B}=-\eth^2\bar\sigma_B-\sigma_B\dot{\bar\sigma}_B+\int_{-\infty}^u \dot\sigma_B \dot{\bar\sigma}_B \mbox{d}\mu
$$
where $\eth$ denote the partial covariant derivative with respect to $\z$. Inserting this in equation (\ref{eq:psi2}) gives

 \be
\label{eq:Psi2}
\Psi_2=-\eth^2\bar\sigma_B-\sigma_B\dot{\bar\sigma}_B-2\eth Z \eth {\dot{\bar\sigma}}_B-(\eth Z)^2 \ddot{\bar\sigma}_B+\int_{-\infty}^Z \dot\sigma_B \dot{\bar\sigma}_B \mbox{d}\mu.
\ee
On the other hand,  Sach's Theorem gives
\be
\label{eq:sach}
\ethbar^2\eth^2 Z=\ethbar^2_T\sigma_B-\ethbar^2\sigma_Z, 
\ee
with $\ethbar_T$ the total derivative, given by 
\be
\label{eq:dertot}
\begin{array}{ll}
\ethbar_T^2 \sigma_B(Z,\zeta,\bar\zeta)&=(\ethbar+\ethbar Z \partial_u)\;(\ethbar+\ethbar Z \partial_u) \sigma_B,\\
&=\ethbar^2 \sigma_B+ \ethbar^2 Z \dot{\sigma}_B+2 \ethbar Z \ethbar\dot{\sigma}_B+(\ethbar Z)^2 \ddot{\sigma}_B\\
&=\ethbar^2\sigma_B+ \bar \sigma_B \dot{\sigma}_B - \bar \sigma_Z \dot{\sigma}_B+2 \ethbar Z \ethbar\dot{\sigma}_B+(\ethbar Z)^2 \ddot{\sigma}_B,
\end{array}
\ee
where Sach's Theorem has been used on the last equality. Inserting (\ref{eq:dertot}) back on (\ref{eq:sach}) and rearranging terms gives
\be 
\label{eq:sach1}
\ethbar^2\sigma_Z + \bar \sigma_Z \dot{\sigma}_B = \ethbar^2\sigma_B+ \bar \sigma_B \dot{\sigma}_B+2 \ethbar Z \ethbar\dot{\sigma}_B+(\ethbar Z)^2 \ddot{\sigma}_B-\ethbar^2\eth^2 Z. 
\ee
Consider now the following combination
$$
- \Psi_2+\ethbar^2 \sigma_Z+\bar \sigma_Z \dot{\sigma}_B,
$$
using (\ref{eq:Psi2}) and (\ref{eq:sach1}) gives
\bea
\ethbar^2\eth^2 Z - \Psi_2+\ethbar^2 \sigma_Z+\sigma_Z\dot{\bar\sigma}_B&=&\left(\eth^2\bar\sigma_B+\sigma_B\dot{\bar\sigma}_B+2 \eth Z \eth\dot{\bar\sigma}_B+(\eth Z)^2 \ddot{\bar\sigma}_B+c.c.\right)+\nn\\
&&-\int_{-\infty}^Z \dot\sigma_B \dot{\bar\sigma}_B \mbox{d} \mu.
\label{eq:freedata}
\eea
The terms on the r.h.s. of (\ref{eq:freedata}) are the boundary terms of eq. (\ref{eq:asymp}), i.e., they vanish when applying $\partial_1$ on this equation. This completes the proof.$\Box$
\end{proof}

%%%%%%%%%%%%%%%%%%%%%%%%%%%%%%%%%%%%%
\section{The Metricity Conditions}
%%%%%%%%%%%%%%%%%%%%%%%%%%%%%%%%%%%%%

In this section we explore possible generalizations of the l.c.cut equations. The idea is to keep the kinematical arena provided by $\scri$ and
replace (\ref{eq:basymt}) with a simpler equation  for the
2-surfaces.  

The main motivation for such a generalization is to obtain a field equation that is a good approximation of the conformal Einstein equations. A related motivation is to be able to obtain analytic solutions and a complete knowledge of the solution space.

It is clear that the surfaces satisfying the new equations
would not yield conformal vacuum metrics but if they, on the other
hand, satisfy \mII they would be characteristic surfaces of an
underlying metric of the space-time with a given dynamical evolution
due to an effective stress-energy tensor.  In particular, we examine
two field equations for cuts, one of them appearing  in the
literature and the other one provided by (\ref{eq:basymt}) with ${\cal
S}=0$. 

Before considering the possible generalizations of (\ref{eq:basymt}) we analyse the linearized version of this equation, namely,

\be
\label{eq:linearZ}
\eth^2\ethbar^2 Z^{(1)}(x^a,\zeta,\bar\zeta)=\eth^2\bar\sigma(x^a l_a,\zeta,\bar\zeta)+\ethbar^2\sigma(x^a l_a,\zeta,\bar\zeta),
\ee
with $x^al_a$ a cut for Minkowski space.

The regular solution to this equation can be written as

$$
Z^{(1)}(x^a,\zeta,\bar\zeta)=\int_{S^2}\left(\ethbar'^2 G_{0 0'}(\zeta,\zeta')\sigma(x^a,\zeta',\bar\zeta')+\eth'^2G_{0 0'}(\zeta,\zeta') \bar\sigma(x^a,\zeta',\bar\zeta')\right)\mbox{d}S'^2,
$$
 where the $G_{0 0'}(\zeta,\zeta')$ is the corresponding Green function on the sphere and the ``prime'' implies dependence in the integrations variables  $(\zeta',\bar\zeta')$.
Applying
$\eth^2$ to the last  equation  we obtain the following expression for $\Ld$.

\be
\Ld(x^a,\zeta,\bar\zeta)=\sigma(x^al_a,\zeta,\bar\zeta)+\int_{S^2}\eth^2G_{0 2'} \; \bar\sigma'\mbox{d}S'^2,
\label{eq:Ld}
\ee
where $G_{0 2'} \equiv \eth'^2G_{0 0'}$.

The linearized metricity condition \mII reads
\be
\label{eq:linII}
\mbox{\mII}\qquad \Ldm-{1\over 2}\eth\Ldo=0.
\ee
Thus, inserting equation (\ref{eq:Ld}) in \mII yields after some manipulations   \[
\Ldm-{1\over 2}\eth\Ldo=-{1\over 2} \int_{S^2}\dot{\bar\sigma'}\left(3\eth^2G_{0 2'}m^a l'_a+\eth^3G_{0 2'}l^a  l'_a\right) \mbox{d}S'^2=0.
\]

We have thus shown that the solution to the linearized field equations (\ref{eq:linearZ}) are the null surfaces of metric $g_{ab}(x^a)$ with
coefficients given by (\ref{eq:metric}). Although not needed for the present discussion, we can also show that $W$ vanishes to first order. A straightforward calculation gives

\[
W=\Ldp+
{1\over 2}\ethbar\Ldo
=-{1\over 2}\int_{S^2}\dot{\bar\sigma'}\left(\ethbar^2G_{0 2'}{\bar m}^a  l'_a-\ethbar\eth^2G_{0 2'}l^a l'_a\right) \mbox{d}S'^2=0.
\]

Thus, the solutions to (\ref{eq:linearZ}) automatically satisfy the metricity conditions \mI and \mII to first order.

We now consider possible generalizations of (\ref{eq:basymt}) that agree with (\ref{eq:linearZ}) to first order.

In a recent paper\cite{mason}, L. Mason conjectured that the following equation,

\be
\label{eq:mason}
\eth^2\ethbar^2 Z=\eth^2\bar\sigma(Z,\zeta,\bar\zeta)+\ethbar^2\sigma(Z,\zeta,\bar\zeta),
\ee
should lead to a generalization of the Bach equation for the l.c. cuts.
Mason also pointed out that the solutions to this equation could yield a
metric of conformal Finsler type, i.e., would not be a conformal metric
for the space-time. Since (\ref{eq:mason}) agrees with the
linearized version of (\ref{eq:basymt}) we should test if \mII is satisfied to second order. We thus consider
$$
\eth^2\ethbar^2 Z^{(2)}(x^a,\zeta,\bar\zeta)=\eth^2\bar\sigma(Z^{(1)},\zeta,\bar\zeta)+\ethbar^2\sigma(Z^{(1)},\zeta,\bar\zeta),
$$
which gives
$$
\Ld(x^a,\zeta,\bar\zeta)=\sigma(Z^{(1)},\zeta\bar\zeta)+
\int_{S^2}\eth^2G_{0 2'} \; \bar\sigma(Z^{(1)},\zeta',\bar\zeta')\mbox{d}S'^2.
$$
Since the second order metricity condition \mII for asymptotically flat spacetimes is identical to (\ref{eq:linII}), we insert the expression for $\Ld$ in this equation obtaining

\begin{eqnarray*}
\Ldm-{1\over 2}\eth\Ldo&=&-{1\over 2}
\int\int_{S^2}\mbox{d}S'^2\mbox{d}S''^2
\dot{\bar\sigma'}\frac{(l^{[a} m^{b]}l'_{[a} m'_{b]})^2}{(l^{c}l'_{c})^4}
l^{[a} m^{b]}l'_{[a} l''_{b]}(G_{0 2''}\sigma'' + G_{0 -2''}\bar\sigma'')\\
&\neq &0.
\end{eqnarray*}
where we have used the explicit form of the Green function and $Z^{(1)}$ to obtain the above expression.

Thus, the solutions to (\ref{eq:mason}) are not characteristic surfaces of an underlying conformal metric. Note that the free data of (\ref{eq:mason}) is given on the cut and thus, the solutions of this equation satisfy Huygens' principle. On the other hand the solutions of (\ref{eq:basymt}) are manifestly non-Huygens. It was thought that replacing the r.h.s. of (\ref{eq:mason}) with the free data obtained in the previous section, namely,

$$
\left(\eth^2\bar\sigma_B+\sigma_B\dot{\bar\sigma}_B+2 \eth Z \eth\dot{\bar\sigma}_B+(\eth Z)^2 \ddot{\bar\sigma}_B+c.c.\right)-\int_{-\infty}^Z \dot\sigma_B \dot{\bar\sigma}_B \mbox{d} \mu.
$$
could improve the situation. An explicit calculation shows that doing this replacement also fails to satisfy (\ref{eq:linII}). 

In both cases the solutions to the field equations would not be null
surfaces of any metric since equation \mII is not satisfied. It is
worth remarking that although the metricity condition \mII is
kinematical in nature it becomes a useful tool to check if solutions to
generalizations for the l.c. cut equations are also characteristic
surfaces.

%%%%%%%%%%%%%%%%%%%%%%%%%%%%%%%%%%%%%
\section{Conclusions}
%%%%%%%%%%%%%%%%%%%%%%%%%%%%%%%%%%%%%

We have shown that the dynamics of the conformal structure can be
written as a single equation for the function $Z$. We proved that this
equation is equivalent to the vanishing of the Bach tensor and wrote
its explicit form to second order in a perturbation expansion. Using
appropriate asymptotic boundary conditions, we obtained the field
equations that determine the global null surfaces for asymptotically
flat space-times. However, it follows from (E) that, once $Z$ is given,
the conformal factor can be explicitly written as a functional of $Z$.
Thus, the field equations for $Z$ determine not only the conformal
structure but the full geometry of the Einstein spaces. We have also
shown that metricity condition \mII becomes an important tool to decide
if generalizations of the field equations of the l.c. cuts also yield a
conformal structure on the manifold.

One of the main applications of the field equations for the l.c. cuts
could be to test the formation of singularities on radiative
space-times. Assume one gives regular initial data $\sigma_B$ at
$\scri^-$ representing incoming gravitational waves. A very important
question, for which there is no answer yet, is to determine whether or
not singularities will be developed in the future. It is generally
accepted that if the data is ``strong enough'' the self interaction of
the gravitational waves will develop a singularity. It is also accepted
that if the data is weak, the waves will scatter into the future
without forming a singularity. The l.c. cuts, on the other hand, can be
used as a tool to detect singularities since their index number is
equal to 1 (topological spheres) if the space-time is regular and 0
(open surfaces) if it has a singularity\cite{kozaleco}. Thus, if we
could show that for a given data $\sigma_B$  the solution space of
(\ref{eq:basymt}) contains a subspace with vanishing index number, we
would prove that the space-time has developed a singularity. Note also
that the conformal factor $\Omega$ plays no role in this discussion.

Another possible line of research would be to obtain generalizations of
the l.c. cut equations that satisfy the metricity conditions or, as
pointed out by Mason, to see if the generalized equations are good
approximations to general relativity for weak fields.

Future work will address the above issues.

\vspace{1cm}

\noindent{\bf Acknowledgments}:

This research has been partially supported by CONICET, CONICOR and STINT.

\newpage
\appendix
%%%%%%%%%%%%%%%%%%%%%%%%%%%%%%%%%%%%%%%%%%%%%%%%%%%%%%%
\section{Proof of Proposition IV.1} \label{Proof1}
%%%%%%%%%%%%%%%%%%%%%%%%%%%%%%%%%%%%%%%%%%%%%%%%%%%%%%%
\begin{proof}
Using commutators to first order in $\Ld$, i. e. $f_i=\partial_i\ethbar\Ld$.
and  applying  $[\partial_1^2,\eth]$ on $\Omega$ gives,  
\bea
[\partial_1^2,\eth]\Omega&=&\partial_1^2\eth\Omega-\eth\partial_1^2\Omega\nn\\
&=&\frac{1}{2}\partial_1^2({W\Omega}).
\label{eq:def11}
\eea
where (E') and \mI have been used in the last equality. On the other hand, using  (\ref{com4}) we compute
\bea
[\partial_1^2,\eth]&=&\partial_1[\partial_1,\eth]+[\partial_1,\eth]\partial_1\nn\\
&=&2\delta_1\partial_1+\eta_1.
\label{eq:id11}
\eea
Therefore (\ref{eq:def11}) and (\ref{eq:id11}), and the first commutation relation given in (\ref{eq:com}), yield

$$
\partial_-\partial_1\Omega+\partial_1 \Ld  \partial_1\partial_+\Omega={1\over 4}\partial_1^2(W\Omega)-{1\over 2}\eta_1\Omega.
$$
In a similar way, we obtain the complex conjugate equation by  applying $[\partial_1^2,\ethbar]$ on $\Omega$. From  these two equations we conclude that $\partial_{\pm}\partial_1\Omega={\cal O}(\Ld)$, thus  at first order in $\Ld$, we can write
\bea
\partial_-\partial_1\Omega&=&{1\over 4}\partial_1^2(W\Omega)-{1\over 2}\eta_1\Omega,\label{eq:d1p}\\
\partial_+\partial_1\Omega&=&{1\over 4}\partial_1^2(\overline{W}\Omega)-{1\over 2}\bar{\eta}_1\Omega.\nn
\eea
Since  these two equations are new differential equations for $\Omega$, they have to be incorporated  to the system (\ref{eq:system1}). Therefore we have to find the integrability conditions for this larger system.
In order to do this we calculate  commutators identities similar  to (\ref{eq:id11}), namely

\be
\label{eq:idij}
\delta_i\partial_j +\delta_j\partial_j=\partial_i\partial_j\eth-[\partial_i,\delta_j]-\eth\partial_i\partial_j.
\ee
We now apply the commutators $[\partial_-\partial_1,\eth]$, $[\partial_+\partial_1,\eth]$ and their complex conjugate on $\Omega$. 
Identity (\ref{eq:idij}), and  equations \mI  and ($\bar m_I$) allows us to calculate
\bea
\partial_-^2\Omega+\partial_1 \Ld \partial_-\partial_+\Omega&=&{1\over 2}\partial_1\partial_-(W\Omega)-[\partial_1,\delta_-]\Omega-\eth\partial_-\partial_1\Omega,\nn\\
\partial_0\partial_1\Omega+\partial_+\partial_-\Omega&=&{1\over 2}\partial_1\partial_+(W\Omega)-[\partial_1,\delta_+]\Omega-\eth\partial_+\partial_1\Omega,\label{eq:dij}\\
\partial_+^2\Omega+\partial_1 \tLd \partial_-\partial_+\Omega&=&{1\over 2}\partial_1\partial_+(\overline{W}\Omega)-[\partial_1,\bar\delta_+]\Omega-\ethbar\partial_+\partial_1\Omega,\nn\\
\partial_0\partial_1\Omega+\partial_+\partial_-\Omega&=&{1\over 2}\partial_1\partial_-(\overline{W}\Omega)-[\partial_1,\bar\delta_-]\Omega-\ethbar\partial_-\partial_1\Omega.\nn
\eea

Note that using the definition of $W$, one can show that the second and the fourth equations are the same, hence the  commutators  $[\partial_i\partial_j,\eth]$ and $[\partial_i\partial_j,\ethbar]$ shall give us less than twenty algebraic linear independent equations with respect to the variables $\partial_i\partial_j\Omega$.

Once again eq.  (\ref{eq:dij}) are new differential equations for $\Omega$ that have to be incorporated to our system. 
Therefore in order to obtain the integrability conditions of the system  (\ref{eq:system1}) we  shall apply on $\Omega$  the commutators $[\partial_i\partial_j,\eth]$, $[\partial_i\partial_j,\ethbar]$ and $[\partial_i\partial_j,\partial_k]$, for $i, j, k=0,+,-,1$.

By inspection, we conclude that if \mI and \mII are satisfied, the system of equations  originated by the commutators $[\partial_i\partial_j,\eth]$ and $[\partial_i\partial_j,\ethbar]$  yield eight linear independent differential equations for $\Omega$, namely
\be
\label{eq:system2}
\left .\begin{array}{rl}
\partial_-\partial_1\Omega&={1\over 4}\partial_1^2(W\Omega)-{1\over 2}\eta_1\Omega,\\
&\\
\partial_-^2\Omega+\partial_1 \Ld \partial_-\partial_+\Omega&={1\over 2}\partial_1\partial_-(W\Omega)-[\partial_1,\delta_-]\Omega-\eth\partial_-\partial_1\Omega,\\
&\\
\partial_0\partial_1\Omega+\partial_+\partial_-\Omega&={1\over 2}\partial_1\partial_+(W\Omega)-[\partial_1,\delta_+]\Omega-\eth\partial_+\partial_1\Omega,\\
&\\
\partial_0\partial_+\Omega+{1\over 2}\eth \partial_1 \tLd \partial_+\partial_-\Omega&={1\over 2}\bigg ({ 1\over 2}\partial^2_+(W\Omega)+\partial^2_1(\overline{W}\Omega)
-2\bar \eta_1\Omega-[\partial_+,\delta_+]\Omega+\\
&\\
&\quad-{1\over 2}\eth\partial_+\partial_1
(\overline{W}\Omega)+\eth\bar [\partial_1,\bar\delta_+]\Omega\bigg )+{1\over 2}\eth\ethbar\partial_1\partial_+\Omega,\\
&\\
\partial_0^2\Omega-\bigg(2-f_+-{1\over 2}\eth^2\partial_1 \tLd \bigg)\partial_0\partial_1\Omega&={1\over 2}\partial_+\partial_0(W\Omega)-[\partial_+,\delta_0]\Omega-{ 1\over 2}\eth\bigg ({ 1\over 2}\partial^2_+(W\Omega)+ \\
&\\
&\quad\partial^2_1(\overline{W}\Omega) -2\bar \eta_1\Omega-[\partial_+,\delta_+]\Omega-{1\over 2}\eth\partial_+\partial_1
(\overline{W}\Omega)+\\
&\\
&\quad+\eth\bar [\partial_1,\bar\delta_+]\Omega\bigg )
-{1\over 2}\eth^2\ethbar\partial_1\partial_+\Omega,
\end{array}
\right \}
\ee
and their complex conjugate. Note that the third and the last equations are real.

\begin{remark}
\label {rem2}
Observing this system, we conclude that it can be generated by applying $\eth$ and $\ethbar$ an appropriate number of times on equation (E').
Consequently the equations of this system  are the components of the trace free part of the field equations calculated in the coordinates frame $\theta^a_i$.
Moreover, note that this system has eight equations and nine unknowns, therefore one of them shall be a parameter, for example $\partial_0\partial_1\Omega$. This parameter corresponds to the trace of the field equations.
\end{remark}

It now remains to consider the commutators $[\partial_i\partial_j,\partial_k]$. 
We begin applying $[\partial_1^2,\partial_-]$ on $\Omega$ as follows.
\bea
0\equiv4[\partial_1^2,\partial_-]\Omega&=&4\partial_1(\partial_1\partial_-\Omega)-4\partial_-(\partial_1^2\Omega),\nn\\
&=&\partial_1^3(W\Omega)-2\eta_2\Omega,
\label{eq:d1W}
\eea
where $\eta_2=[\partial_1,\eta_1]$ , and (E') and the first equation of  (\ref{eq:system2}) have been used in the last equality.
 Similarly, we calculate its complex conjugate expression.
\be
\label{eq:bard1W}
0\equiv4[\partial_1^2,\partial_+]\Omega=\partial_1^3(\overline{W}\Omega)-2\bar\eta_2\Omega.
\ee
Note that (\ref{eq:d1W}) and (\ref{eq:bard1W})  only contain first derivatives of $\Omega$ since  all second derivatives can be replaced using (E'). 

We obtain the next integrability condition by applying  the commutator $[\partial_+\partial_1,\partial_-]$ on $\Omega$

\bea
0\equiv4[\partial_+\partial_1,\partial_-]\Omega&=&4\partial_+(\partial_1\partial_-\Omega)-4\partial_-(\partial_+\partial_1\Omega),\nn\\
&=&\partial_1^2\partial_+(W\Omega)-\partial_1^2\partial_-(\overline{W}\Omega)-2\bigg(\partial_+\eta_1\Omega-\partial_-\bar\eta_1\Omega\bigg),
\label{eq:d+1-}
\eea
here we have used the expressions for $\partial_1\partial_\pm \Omega$ given in the system (\ref{eq:system2}). Furthermore we calculate 
\begin{eqnarray*}
0\equiv4[\partial_1^2,\partial_0]\Omega+4[\partial_1\partial_-,\partial_+]\Omega&=&4\partial_1\bigg(\partial_0\partial_1\Omega+\partial_-\partial_+\Omega\bigg)-4\partial_+(\partial_1\partial_-\Omega),\\
&=&\partial_1^2\partial_+(W\Omega)-4\partial_1[\partial_1,\delta_+]\Omega-4\partial_1\eth\partial_+\partial_1\Omega+2\partial_+\eta_1\Omega,
\end{eqnarray*}
the last equations have been obtained using once again the corresponding expression for the second derivatives $\partial_0\partial_1\Omega+\partial_-\partial_+\Omega$ and $\partial_1\partial_-\Omega$   given in (\ref{eq:system2}). Finally, replacing $\partial_1\partial_+\Omega $, using the integrability condition 
(\ref{eq:d+1-}) and commutating a couple of times, we obtain
\be
\label{eq:d110}
0\equiv4[\partial_1^2,\partial_0]\Omega+4[\partial_1\partial_-,\partial_+]\Omega=-\eth\partial_1^3(\overline {W}\Omega)+2\eth\bar\eta_2\Omega.
\ee
Observe that the integrability condition (\ref{eq:d110}) can be obtained equivalently by applying $\eth$ on equation (\ref{eq:d1W}) thus, as in the case of the field equations, we generate linear combinations of the integrability condition  corresponding to $[\partial_i\partial_j,\partial_k]\Omega$ by applying $\eth$ and $\ethbar$ an appropriate number of times on  equation (\ref{eq:d1W}).

Summarizing,  the original system has become even  larger and it remains to find equations independent of $\Omega$ as integrability conditions. One of these conditions can be obtained by applying  $\partial_1 $ on  equation (\ref{eq:d110})  and using (\ref{com4}) as follows.

\bea
4\partial_1\eth[\partial_1^2,\partial_+]\Omega&=&\partial_1\eth\left(\partial_1^3(\overline{W}\Omega)-2\bar \eta_2\Omega\right),\nn\\
&=&\partial_1^4\eth(\overline {W}\Omega)-\partial_1[\partial_1^3,\eth](\overline {W}\Omega)-2\partial_1\eth\bar \eta_2\Omega,\nn\\
&=& \partial_1^4\eth(\overline {W}\Omega)-  3\partial_-\partial_1^3(\overline{W}\Omega)-2\partial_1\eth\bar\eta_2\Omega,\nn\\
&=&\partial_1^4\eth(\overline {W}\Omega)-  3\partial_-(4[\partial_1^2,\partial_+]\Omega+2\bar\eta_2\Omega)-2\partial_1\eth\bar\eta_2\Omega,
\label{eq:Bacha}
\eea
in the last equation we have used the integrability condition  (\ref{eq:bard1W}). Consequently we write

\bea
0\equiv4\partial_1\eth[\partial_1^2,\partial_+]\Omega+12\partial_-[\partial_1^2,\partial_+]\Omega&=&\partial_1^4\eth(\overline {W}\Omega)-6\partial_-\bar\eta_2\Omega-2\partial_1\eth\bar\eta_2\Omega\label{eq:Bach11}\\
&=&\Omega\partial_1^4\eth\overline {W}+4 \partial_1^3\eth\overline{W}\partial_1\Omega-6[\partial_-,\bar\eta_2]\Omega-2[\partial_1\eth,\bar\eta_2]\Omega,\nn
\eea
here we have used the system (\ref{eq:system2}), and equations (E') and \mI. Note that this equation contains  only first derivatives of $\Omega$. Using now the expression for $W$ and calculating the commutators $[\partial_+,\bar\eta_2]$ and $[\partial_1\eth,\bar\eta_2]$, we find that the coefficients enclosed to $\partial_i\Omega$ are zero  and as consequence the equation (\ref{eq:Bach11})  becomes
$$
\partial_1^4\eth\overline{W}=0.
$$

Note that we have a similar equation corresponding to the complex conjugate commutators, i.e $ \partial_1^4\ethbar W=0$.

Since at first order $\eth \overline{W}=\ethbar W$, we get the same integrability condition. Using again the definition of $W$, we obtain a differential  equation for $\Ld$, namely
$$
\partial^5_1\eth^2\tLd=0.
$$
This completes the proof.$\Box$
\end{proof}

%%%%%%%%%%%%%%%%%%%%%%%%%%%%%%%%%%%%%%%%
\section{Proof of Proposition IV.2}
%%%%%%%%%%%%%%%%%%%%%%%%%%%%%%%%%%%%%%%%%
\label{Proof2}
\begin{proof}
Following the same approach that leads to (\ref{eq:system2}) but using the full commutators gives the following system of equations

\be
\label{eq:system3}
\left .
\begin{array}{rl}
\partial_-\partial_1\Omega+\partial_1 \Ld  \partial_1\partial_+\Omega&=\displaystyle{{1\over 4}}\partial_1^2(W\Omega)-\displaystyle{{1\over 2}}\eta_1\Omega-\displaystyle{{1\over 2}}\eth(Q \Omega) - f_1 Q \Omega,\\
&\\
\partial^2_-\Omega+\partial_1 \Ld  \partial_-\partial_+\Omega&=\displaystyle{{1\over 2}}\partial_1\partial_-(W\Omega)-[\delta_-,\eth]\Omega-f_-Q\Omega-\partial_- \Ld \partial_+\partial_1\Omega+\\ 
%&\\ 
&\quad-f_1\partial_-\partial_1\Omega-\eth(\partial_1\partial_-\Omega),\\
&\\
\partial_0\partial_1\Omega+q\partial_+\partial_-\Omega&=\displaystyle{{1\over 2}}\partial_+\partial_1(\Omega W)-[\delta_+,\eth]\Omega-f_+ Q \Omega-(f_1+\partial_+\Ld)\partial_1\partial_+\Omega+\\
%&\\
&\quad-\eth \partial_1\partial_+\Omega+\partial_1 \Ld \bigg(\displaystyle{{1\over 2}}\partial_1\partial_-(W\Omega)-[\delta_-,\eth]\Omega+\\
%&\\
&\quad-f_-Q\Omega-\partial_- \Ld \partial_+\partial_1\Omega-f_1\partial_-\partial_1\Omega-\eth(\partial_1\partial_-\Omega)\bigg)
\end{array}
\right\}
\ee 
and their complex conjugate.
Since the last  equation is real, this system  contains only five linear independent equations.

The remaining  equations for $\partial_0^2\Omega$ and $\partial_+\partial_0\Omega$ and its complex conjugate are obtained by applying $\eth$ and $\eth^2$ on the last equation of the system  (\ref{eq:system3}).

We now consider the commutators $[\partial_i\partial_j,\partial_k]\Omega$. Taking the first equation of (\ref{eq:system3}) and commuting with $\partial_1$ gives

\be
\label{eq:C11+}
\begin{array}{ll}
[\partial_1^2,\partial_-]\Omega+\partial_1 \Ld [\partial_1^2,\partial_+]\Omega&=[\partial_1^2,\delta_1]\Omega-f_1[\partial_1^2,\partial_1]\Omega-2\eta_1\partial_1\Omega-\eta_2\Omega\\
&\\
&=\partial_1(\partial_1\delta_1\Omega)-\delta_1(Q\Omega)-2\eta_1\partial_1\Omega-\eta_2\Omega\\
&\\
&={1\over 4}\partial_1^3(W\Omega)-{1\over 2}\partial_1\eth(Q\Omega)-{3\over 2}\eta_1\partial_1\Omega-{1\over 2}\eta_2\Omega -\delta_1(Q\Omega),
\end{array}
\ee

Thus, algebraically solving for (\ref{eq:C11+}) and its c.c. yields

\be
\label{C11+}
0\equiv 4[\partial_1^2,\partial_+]\Omega=c_{11+},
\ee
where 
\bea
c_{11+}&\equiv&{1 \over q}\left(\partial_1^3(\overline{W}\Omega)- \partial_1 \tLd  \partial_1^3(W\Omega)\right )-{6\over q}\left(\bar\eta_1\partial_1\Omega-\partial_1 \tLd  \eta_1\partial_1\Omega\right)+\label{def:C11+}\\
&&-{2\over q}\left(\bar \eta_2\Omega-\partial_1 \tLd  \eta_2\Omega\right )-{2\over q}\left(\partial_1\ethbar(Q\Omega)- \partial_1 \tLd  \partial_1\eth(Q\Omega)\right)\nn\\
&&-{2\over q}\left(\bar\delta_1(Q\Omega)-\partial_1 \tLd  \delta_1(Q\Omega)\right)\nn.
\eea

As in the first order calculation we obtain new integrability conditions calculating 
$[\partial_1^2,\partial_+]\Omega$ and $[\partial_1\partial_+,\partial_-]\Omega$.

In order to obtain the integrability condition corresponding to
equation (\ref{eq:d110}), we  follow  the same procedure  as in the
first order calculation giving

\begin{eqnarray*}
0&=&4[\partial_1^2,\partial_0]\Omega+4q[\partial_1\partial_-,\partial_+]\Omega,\\
&=&4\bigg(\partial_1(\partial_0\partial_1\Omega+q\partial_+\partial_-\Omega)-\partial_0(\partial_1^2\Omega)-q\partial_+(\partial_1\partial_-\Omega)-\partial_1q(\partial_-\partial_+\Omega)\bigg).
\end{eqnarray*}
Using the corresponding equations of the system (\ref{eq:system3}),  and (E), and the integrability condition given by $[\partial_1\partial_+,\partial_-]\Omega=0$, we obtain 

\bea
0&\equiv& 4[\partial_1,\partial_0]\Omega+4q[\partial_1\partial_-,\partial_+]\Omega\nn\\
&=&-\left(\eth c_{11+}+Fc_{11+}\right),
\label{eq:M110}
\eea
 where 
\be
F=2f_1+\partial_+\Ld-{1\over 2}(\ethbar \partial_1 \Ld +\overline{W}\partial_1 \Ld ).
\label{eq:F}
\ee
Note that equation (\ref{eq:M110}) can be obtained by applying the operator $(\eth+F)$  on the integrability condition (\ref{C11+}).

The integrability condition independent of $\partial_i\Omega $ corresponding to equation (\ref{eq:Bach11}) can essentially be calculated following the same procedure as in the first order calculation.

 Namely, we take $\partial_1$ to equation (\ref{eq:M110}) and  commutate $\partial_1^3$ with $\eth$. On the other hand  we  apply the operator $ (\partial_-+{1\over 2}(\ethbar \partial_1 \Ld +\overline{W} \partial_1 \Ld )\partial_1+\partial_1 \Ld  \partial_+ ) $ on  equation (\ref{C11+}) and replace this result in the previous equation  or equivalently
\begin{eqnarray}
0&\equiv &4\partial_1\left(\eth[\partial_1^2,\partial_+]\Omega+F[\partial_1^2,\partial_+]\Omega\right)+12\left(\partial_-+{1\over 2}(\ethbar \partial_1 \Ld +\overline{W} \partial_1 \Ld )\partial_1+\partial_1 \Ld  \partial_+\right)[\partial_1^2,\partial_+]\Omega
\nn\\
&=&\partial_1\left(\eth c_{11+}+Fc_{11+}\right)+3\left(\partial_-+{1\over 2}(\ethbar \partial_1 \Ld +\overline{W} \partial_1 \Ld )\partial_1+\partial_1 \Ld  \partial_+\right)c
_{11+}.
\label{B11complex}
\end{eqnarray}
When this expression is written explicitly using the previous equations,  the r.h.s. of this integrability condition has no second derivatives of $\Omega$ except for a term with $\partial_0\partial_1\Omega$. The coefficient of this term is purely imaginary, thus the integrability condition is given by the real part of (\ref{B11complex}). 

We also claim that the real part of eq. (\ref{B11complex}) does not
contain first derivatives of $\Omega$ . To prove this, instead of doing a
straightforward calculation,  we follow a tensorial approach establishing
a correspondence between the integrability condition
(\ref{B11complex})  and the coordinates components of the Bach tensor (for details see \cite{teretesis}).$\Box$
\end{proof}

\end{document}